# AIMS-EREA - A framework for "AI-accelerated Innovation of Materials for Sustainability - for Environmental Remediation and Energy Applications"

Sudarson Roy Pratihar *
Data Science
SymphonyAI
Bangalore, India
sudarson@symphonyai.com

Deepesh Pai
Data Science
SymphonyAI
Bangalore, India
deepesh.pai@symphonyai.com

Manaswita Nag
Materials Science
Indian Insitiue of Science Alumni
Bangalore, India
drmanaswitanag@gmail.com

*Abstract*— Many environmental remediation and energy applications (conversion and storage) for sustainability need design and development of green novel materials. Discovery processes of such novel materials are time taking and cumbersome due to large number of possible combinations and permutations of materials structures. Often theoretical studies based on Density Functional Theory (DFT) and other theories, coupled with Simulations are conducted to narrow down sample space of candidate materials, before conducting laboratory-based synthesis and analytical process. With the emergence of artificial intelligence (AI), AI techniques are being tried in this process too to ease out simulation time and cost. However tremendous values of previously published research from various parts of the world are still left as labor-intensive manual effort and discretion of individual researcher and prone to human omissions.  AIMS-EREA is our novel framework to blend best of breed of Material Science theory with power of Generative AI to give best impact and smooth and quickest discovery of material for sustainability. This also helps to eliminate the possibility of production of hazardous residues and bye-products of the reactions. AIMS-EREA uses all available resources - Predictive and Analytical AI on large collection of chemical databases along with automated intelligent assimilation of deep materials knowledge from previously published research works through Generative AI. We demonstrate use of our own novel framework with an example, how this framework can be successfully applied to achieve desired success in development of thermoelectric material for waste heat conversion.

## I. Introduction

After a few decades of industrial revolutions, a growing concern of environmental impacts and need of sustainable development have been a strong realization. Every effort is made for *"meeting the needs of the present without compromising the ability of future generations to meet their own needs"*.

Development of novel materials plays a pivotal role in addressing sustainable development and environmental challenges at large scale. Energy applications (conversion of waste heat, renewable energy, storage of energy etc.) and environmental remediation (Carbon di oxide reduction, water treatment etc.) are two critical areas where accelerated development of novel materials is much needed, since present commercially available materials or devices are sub-optimal with respect to efficiency, toxicity, and cost. [1]

As we progress from height of Artificial Intelligences (AI) to realm of Artificial General Intelligence (AGI), material scientists aspire to revolutionize intricate process of discovery and development of novel environment friendly materials to promote sustainability.

However, this endeavor needs to address a complex set of problems. Scientific knowledge is growing at tremendous rate (~500 papers / day) [2] and numerous sets of application programming interfaces (API) and databases of scientific experiments and theoretical studies are growing too [3]. Since target objective varies for materials development in each problem area, neither traditional software programs nor one single finest AI technique can holistically address this.  Multi-disciplinary knowledge in various branches of science and technology such as materials science, thermodynamics, electronics, computer science, data science, machine learning, statistics and labor-intensive programming makes this even harder. This necessitates a framework which can plug-in various tools and techniques from arsenal of traditional and AI tools and dynamically orchestrate them to achieve the target.

In response to these challenges, we present the "**AI**-accelerated **I**nnovation of **M**aterials for **S**ustainability in **E**nvironmental **R**emediation and **E**nergy **A**pplications" framework, abbreviated as AIMS-EREA, in short AIMS (Figure 1). AIMS is a framework with a materials science reasoning brain and a set of intelligent tools. AIMS allows us to easily plug in additional tools or new intelligence. Recent developments in large language models (LLMs) motivated us to build this core reasoning backbone using LLM.  AIMS "brain" breaks a complex materials science problem and coordinates the tools to solve it. This integrated approach not only accelerates the process, but also substantially reduces the possibility of generating hazardous residues and by-products during materials development. In this paper, we will go through the features and architecture of AIMS-EREA and demonstrate the power of AIMS-EREA through an accelerated material discovery process of thermoelectric material for waste heat conversion.

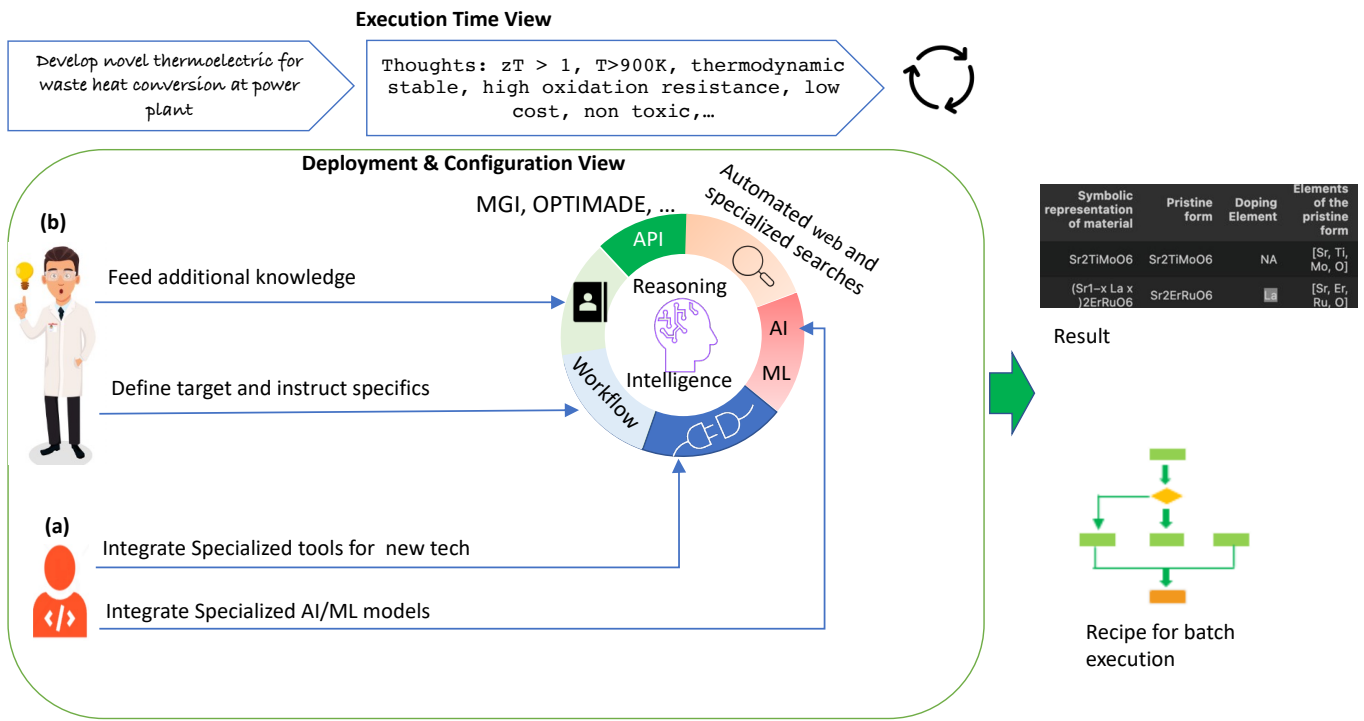

Fig. 1. Schematic presentation of AIMS-EREA showing 2 views. **Deployment and Configuration view**: (a) Developer has ability to plug-in new tools and technology and AI/ML models to make it future proof. (b) Expert Chemical scientist can add specific knowledge in terms of selected research papers as well as curated knowledge in natural lanugage (presently in English) to AIMS-EREA. Expert can define target and specific instructions for solving a complex task to workflow module. **Execution time view:** When end user ask to execute a specific task, AIMS-EREA, Reasoning and intelligence module (which is core to the entire system) break it down using implied targets and create a dynamic workflow combining all available tools: viz. search, open material database thru API, AI-ML models for estimating parameters etc. A subset of result is shown for validation along with generated recipe for batch execution of the generated workflow. Upon validation, end user can run AIMS-EREA along with generated recipe for full run at large scale

## II. DESIGN & ARCHITECTURE

### A. Design Considerations, Personas and Features

Purpose of our work is to bring an AI & LLM powered framework (along with key foundational and reusable components) to realize orchestrated power of present and future tools and techniques for accelerated discovery and synthesis of materials for environmental and energy applications. The framework has been designed to serve below two personas:

*1) Materials Scientist:* Needs ability to easily discover key information accurately from till date works, open databases, and APIs (viz. Materials Project and other APIs) for material information. *Discovery* uses multiple search tools (such as google scholar, Scopus) along with curated set of literatures (and specific scientific problem-solving steps) by the researcher. Performing this operation at large scale by materials scientist needs discovery execution at scale to be low cost and simpler. While ensuring accuracy, prevention of hallucinations are key considerations, too.

*2) Developer/Integrator:* To make this future proof, an ease of integration of new tools and techniques are provided to integrate any new search or discovery tool along with integration of Data API, LLM tools, AI models and Simulation models (such as DFT) for expanding capabilities.

### B. Architecture and Modules (Figure 2)

*1) **R**eseasoning and **I**ntelligence **A**gent (RIA):*

This is the core component and *brain* of AIMS-EREA. LLM's capability to reason and breakdown of complex problems is utilized to solve problems for materials scientist. LangChain framework [3] and its agents have been used along with LLM. "Methodology" section discusses this part in detail.

*2) Workflow module*

This module allows users to add standard as well as custom instruction sets to AIMS-EREA for performing specific tasks. Materials scientists can instruct AIMS-EREA to use specific instructions set. This gives the ability to the scientist to follow a particular workflow than using standard path. Alternatively, an agent in AIMS-EREA (RIA) can pick the most suitable instruction set, when not specified. In context learning of LLMs are leveraged to build specific workflow based on the instruction set. [4]

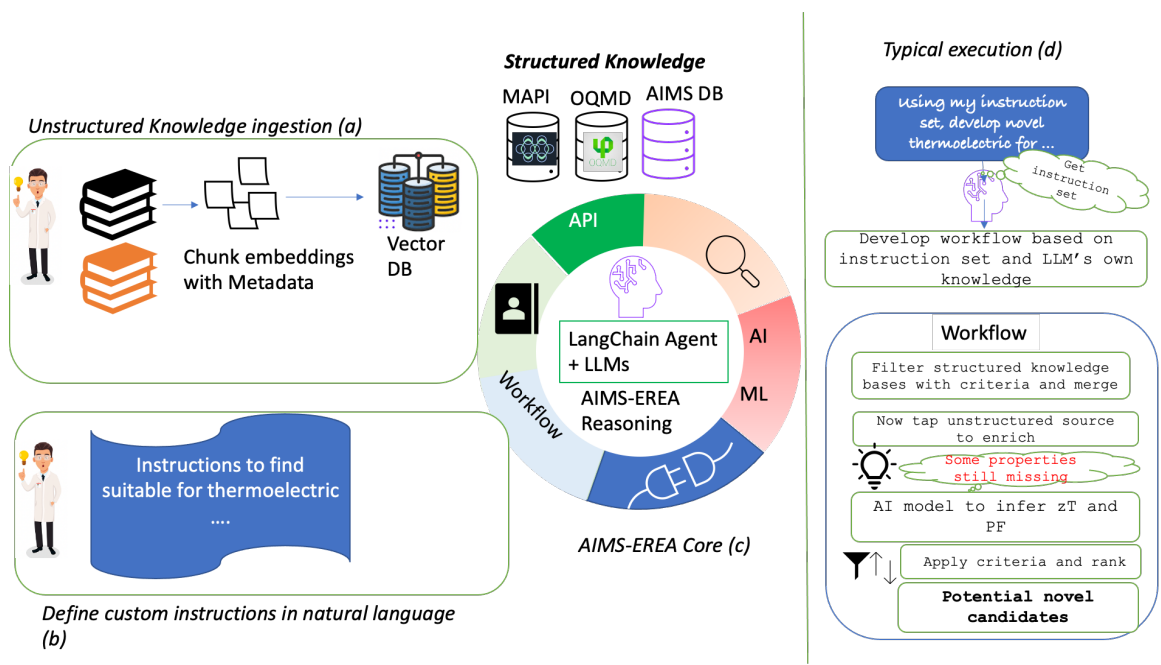

Fig. 2. AIMS-EREA architecture showing (a) Ingestion of knowledge from unstructured sources such as publications, (b) Material Scientist's instruction sets (i.e. prompts) for tasks, (c) AIMS-EREA core module, (d) a typical execution flow inside AIMS-EREA to solve a particular problem

*3) Knowledge Ingestion*

Keeping up-to-date knowledge is a key challenge for AIMS-EREA. Pretrained as well as fine-tuned LLMs suffer from the fact that knowledge base does not get refreshed. We have provided provisions to ingest structured as well as unstructured (text or PDF) knowledge as below.

*a) Structured Knowledge (Through API module)*

API module provides ability to plug-in tools to source structured information from open databases (and APIs) such as Materials Project API (MAPI), Open Quantum Materials Database (OQMD) as well as from internal database of AIMS-EREA. This allows ingestion of external knowledge as well as ingestion of custom knowledge parsed through custom parsers [5]. The module can be extended further by plugging in additional tools for any external API or database.

*b) Unstructured Knowledge (Separate module)*

Publications consists of bulk of latest knowledge in materials sciences. Parsing journals and publications are a labor and programming intensive process.[5]. Here we have provided a LLM powered knowledge ingestion and retrieval mechanism in this module to overcome this issue. Each research paper is split and tokenized into small chunks (Figure 3). Each chunk is then converted into embeddings. Each embedding is also supplemented with meta-data about the chunk such as [doi, type_of_paper, material_type, section, etc.]. Some information such as doi, type of paper is extracted once from the abstract section and then passed on to each chunk.

AIMS-EREA retrieves accurate information efficiently by filtering through metadata and semantic similarity of embeddings of user query and embeddings of chunks.

*4) Search*

This module provides a facility to integrate different search APIs such as google search, Scopus search to get relevant information. Currently google search API has been integrated, but more search tools can be integrated.

*5) AI ML*

Researchers can integrate AI ML models to infer values of properties of materials. AIMS EREA will automatically use these models based on instruction sets in the workflow module.

*6) Pluggability*

This module enables us to integrate more tools and techniques to AIMS-EREA. An immediate use of this is to integrate DFT model built by researchers to predict values for novel materials where not much study has been done. Also, developers and integrators of new tools can use this module to expand capabilities of AIMS-EREA for generic or specific purposes.

*7) Batch vs Inteactive mode*

Discovery of sustainable material (and other tasks associated with it) could be quite complex and time consuming. To make it efficient and at the same time friendly to the researchers, AIMS EREA provides interactive mode as well as batch mode of operation. Interactive mode is meant for accomplishing simple tasks or working on small datasets. To work in batch mode, the user needs to generate workflow file. Workflow files can be created in interactive mode by instructing appropriately. AIMS EREA will create a sample output for the user to verify and on confirmation, a workflow file will be generated.

III. METHODOLOGY

A. *Reasoning and Inteligence*

Commoditization of LLM through proprietary as well as open sources has opened vast opportunity for creating reasoning and intelligent agent. To bring complex problem solving through reasoning has been possible through application of appropriate neuro-symbolic architecture like "Modular Reasoning, Knowledge and Language" (MRKL, pronounced "miracle")[15], Reasoning and Acting (ReAct)[16], Plan and Solve[17], Chain-of-thoughts[18] and tree-of-thoughts[19]. This capability is leveraged to create RIA for AIMS-EREA. Commoditized LLMs such as OpenAI's GPT4 are foundational models and have been trained on various topics including deep science. To train these LLMs at materials science, we need to train them through finetuning or in-context-learning (ICL). Fine tuning is costly and is available on certain models of LLMs. ICL, which is also known as prompt engineering, has been gaining popularity to instruct LLM to work in certain way. We have used prompt engineering for LLM.

LangChain is an opensource framework complements LLM by providing agent incorporating neuro-symbolic architecture. Also, langchain adds tools to LLM to extend its capability beyond its self-contained space. Tools are software program capable to perform a specific task, such as web search, document parsing, invoking API, accessing databases etc.

We have used a combination of OpenAI GPT4 with Langchain to create RIA for AIMS-EREA.

B. *Expanding capability through tools*

Langchain framework allows to expand the capability of LLM by adding external tools. Below are the tools listed as per their placement in AIMS-EREA architecture.

**Tools:**

  *1) Tools for Structured Knowledge:*

Materials Project API tools – A set of tools leveraging mp_client, matminer, automatminer and pymatgen libraries from Materials Project API [6-10] to get material information, theoretical research, and thermodynamic properties. These tool sets include AI ML capabilities of matminer and automatminer as well.

OQMD API tool – Tool to access open quantum mechanics database is based on Restful API of OQMD. [11-12]

  *2) Tools for unstructured data retrieval:*

This is a custom tool to search user query against stored research papers in vector database. This is discussed in detail in methodology for information extraction from unstructured knowledge.

  *3) AI ML tool*

We have built a Light Gradient Boosting Machine (LGBM) [13] regression model and trained it structured data from various sources [5, 6, 7]. Inference pipeline code have been exposed as AI ML regression tool to predict properties of materials.

  *4) Search Tools*

Google Search, Toxic material, and abundance search

  *5) General Tools*

Python REPL Code generator, Pandas[14] tool for dataset processing

C. *Unstructured Knowledge Storage and Retrieval*

As we have touched upon this topic in architecture section, retrieving information from vast collection of knowledge articles is a major task. We have developed hierarchical algorithm to preserve contextual importance of each portion of the document. And then store the semantic representation of same. Retrieval happens based on semantic similarity within that context.

IV. RESULTS AND DISCUSIONS

A. *Human-AI Material Interactive Discovery*

Material discovery is complex multidisciplinary area. A successful collaboration between human researcher and machine can provide better results than each one of them alone. It reduces risks of machine error in interpreting complex and emerging theory. AIMS-EREA's main objective is to break down complex problem of material development (for environmental and energy applications); use various sources of data, API and computational chemistry tools to solve each broken down step; and get the final recommendations. This helps a researcher to focus more on core experimental part, than performing either a lot of manual tasks or a lot of programming to solve each task. AIMS-EREA's core module RIA is tasked to do these.

AIMS-EREA did a thorough job along with explanations when user asked to assist in discovering a thermoelectric material for waste heat conversion. It picked up right instruction set (when provided) and blended that with LLM's own chemistry knowledge and reasoning capability. For example, in case of missing values, RIA tried looking up for AI ML regression module to predict values. RIA also use its best judgement to keep a few materials in recommendations instead of strict rules.

For each such interactive execution (Figure 3), we have verified the steps and reasoning and as well as result at each step. To evaluate accuracy of the final result (Table 1), we actually took data used by each step and manually processed that to arrive at the result. This process involves accessing data from Materials Project, searching Materials Web to get thermodynamic properties, look up at National NIOSH for toxicity, abundance check from Wikipedia. Results matched.

# Chatbot

🤖 Assist in discovery of novel thermo electric materials for waste heat conversion in high temperature. Leverage expert instructions sets available for this purpose

🤖 Looking up ..... Got instructions set for thermoelectric discovery

**Thoughts** : I want to follow the step by step instructions to find thermoelectric material.

**Thoughts**: Material should have band_gap <=3.5. Using MAPI tools to retrive band gap information. Limiting records to 100.

**Action**: Retrieve 100 materials and their atomic and electronic properties

**Thoughts**: Remove toxic material using Toxicity checker tool.

**Action**: Remove materials which are highly toxic. Use Pandas Tool for data querying and procesisng.

**Thoughts**: Find stable materials by looking at eHull. Use pandas tool to List out materials with eHull 0 or negative.

**Action**: Marking those materials as stable where eHull is 0 or negative, but still keeping others for further analysis.

**Thought**: Next available Instructions: abundance >= 0.000001. Use pandas tool to filter less abundant materials

**Action**: Remove materials containing elements < 0.000001

Now I have more finer set of materials. I need to find thermoelectric material from this set.

**Thoughts**: Efficiency of thremoelectric material is found using zT.

**Action**: I need to find zT for each material using knowledge retriever tool and need to sort material it.

**Observation**: I have sorted material based on zT. Found 4 materials with eHull 0 or negative.

**Action**: I need to recommend the best material to user. I need to display top 3 material to user.

**Final Answer**:

```
|            Material             | zT  | Temperature |                                     Remarks                                       |
| Ca2ZrTiO6                       | 4.4 |    500K     | Thermodynamic stability to be verified                                             |
| Sr0.09Ba0.11Yb0.05Co4Sb12       | 1.6 |    800K     | Sb is toxic in nature. So detail study of toxicity in material and synthesis needed |
| n-type nano-structured SiGe     | 1.3 |    700K     | Bulk form this material is widely used. So detail study of novelty needed.         |
| MgTa2O6                         | 1.1 |    1200K    |                                                                                    |
```

Fig. 3. Shows output of one interactive session to discover thermoelectric material for high temperature application. RIA has shown all reasoning and steps followed.

TABLE I. RESULTS OF THERMOELECTRIC MATERIAL DISCOVERY

| Material | $zT \downarrow$ | Temperature (K) | Recommendation |
|---|---|---|---|
| $Ca_2ZrTiO_6$ | 4.4 | 500K | Thermodynamic stability to be checked |
| $Sr_{0.09}Ba_{0.11}Yb_{0.05}Co_4Sb_{12}$ | 1.6 | 800K | Possible toxicity due to Sb |
| n-type nano-structured SiGe | 1.3 | 700K | Bulk form widely used. Novelty to be studied. |
| $MgTa_2O6$ | 1.1 | 1200K | |

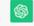

Fig. 4.     Results from ChatGPT as base model

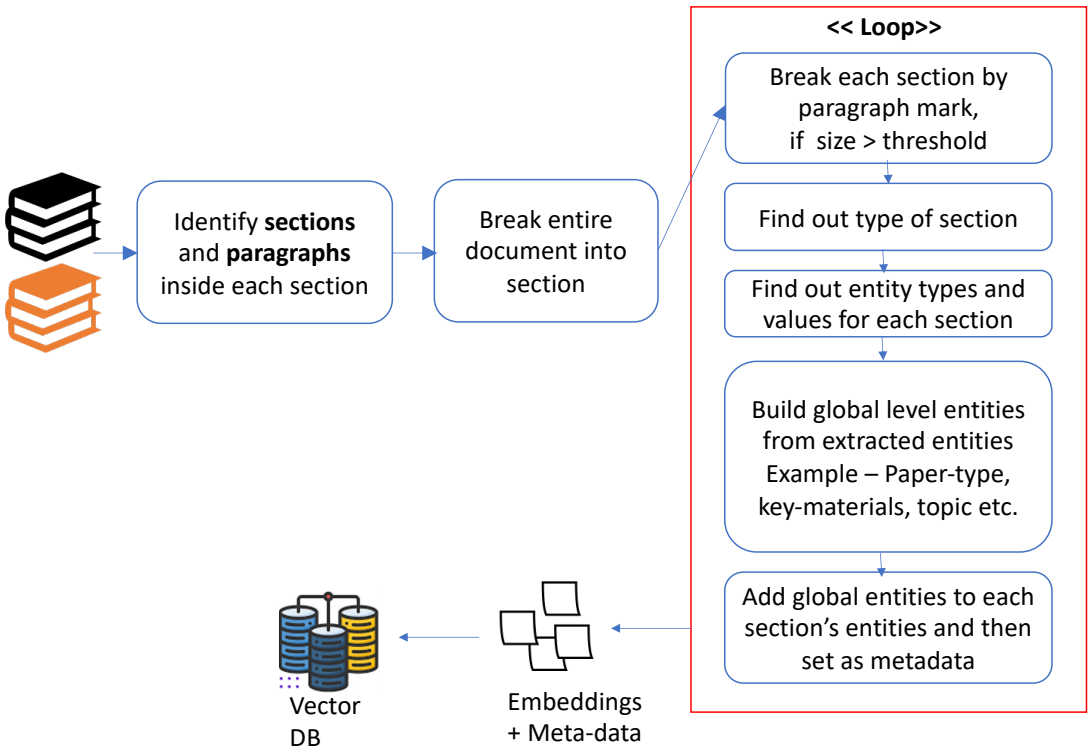

Fig. 5.     Enhanced knowledge retrieval scheme

Having verified our result accuracy, we compared its performance against base model with OpenAI ChatGPT (Figure 4). This shows AIMS-EREA's very own architecture and methodology extends its capabilities deep in material sciences and this is superior to intrinsic knowledge of bare LLM alone.

*B. Chemistry comprehension*

Material science terminologies, formulae and notations are complex and have different meaning than their literary English meaning. is a complex topic. Our evaluation shows more than 99.9% accuracy in understanding language of chemistry and materials science.

*C. Knowledge Retrieval*

For knowledge retrieval from research papers, apart from ability to plug-in in external tools or scripts, AIMS-EREA also provides an optional module. So, we additionally evaluated accuracy of this optional module from research papers in local repository. For this we used 108 papers and manually retrieved 3 parameters for each material – thermal conductivity, zT and Temperature. And matched the same with information retrieved by AIMS-EREA. We found a challenge that parameters values are inter-mingling when multiple materials are mentioned in the same paper as part of literature survey.

To overcome this, we experimented with an enhanced scheme to embed every logical section of the paper [Figure 5]. This made knowledge ingestion to read each section separately and retrieve key information. As an example - "Abstract" section is read to understand nature of the paper and key material that is used there. This approach shown potential to overcome the problem to a great extent – with a precision > 97% and recall > 75%. There are opportunities to further improve these numbers. Apart from manual verification, we also verified materials and extracted parameters against similar published work[5].

## V. CONCLUSION

Our work has demonstrated reasoning ability of LLM to integrate various chemistry knowledge sources and orchestrate multiple APIs to help accelerate material development for energy and environmental application. We also demonstrated blending of expert knowledge with open knowledge databases and vast collection of research papers to solve a complex material discovery problem. This approach also enables material researchers to use capability of LLM within guardrails through expert instructions to avoid hallucinations. Our developed AIMS-EREA research copilot (i.e., an assistant to researcher) has shown the ability to operate in interactive mode and shows potential to operate in batch mode for large scale material research and development in industry.

Although our work has been limited with a set of limited tools for material developments, our results show promising potential to expand AIMS-EREA in various areas of material development, ranging from studies in discovery, synthesis to theoretical analysis. Power of AIMS-EREA comes from orchestration of tools. Extending its arsenal of tools can expand its capabilities many folds. We envisage that AIMS-EREA will serve as foundational framework in areas of materials development.